\documentclass{iaus}
\usepackage{graphicx}
\usepackage{makeidx}
\usepackage{natbib}

\newcommand{\rhk}{$R^\prime_{\rm HK}$}

\newcommand{\logrphk}{$\log R^\prime_{\rm HK}$}

\newcommand{\rx}{$R_{\rm X}$}

\newcommand{\logrx}{$\log R_{\rm X}$}
\newcommand{\loglxlbol}{log($L_{\rm X}$/$L_{bol}$)}

\newcommand{\ro}{$R_o$}

\newcommand{\bvo}{$(B - V)_0$}

\newcommand{\aap}{{\textit A\&A}}

\newcommand{\aj}{{\textit AJ}}
\newcommand{\apj}{{\textit ApJ}}
\newcommand{\apjl}{{\textit ApJ}}
\newcommand{\apjs}{{\textit ApJS}}

\newcommand{\jgr}{{\textit Jnl. Geophys. Res.}}
\newcommand{\mnras}{{\textit MNRAS}}
\newcommand{\nat}{{\textit Nature}}
\newcommand{\pasp}{{\textit PASP}}
\newcommand{\procspie}{{\textit Proc. SPIE}}

\usepackage{index}
\newindex{author}{idx}{ind}{Author Index}
\newindex{object}{odx}{ond}{Object Index}
\newindex{subject}{sdx}{snd}{Subject Index}
\makeindex

\title[Solar-type dwarf ages] {How accurately can we age-date solar-type
dwarfs using activity/rotation diagnostics?}

\author[Eric E. Mamajek]  
{Eric E. Mamajek}
\index[author]{Mamajek, E.E.}

\affiliation{Department of Physics \& Astronomy,\\
University of Rochester, Rochester, NY 14624 USA\\
email: {\tt emamajek@pas.rochester.edu}}

\pubyear{2009}
\volume{258} 
\pagerange{375-382}
\jname{The Ages of Stars}
\editors{E.E. Mamajek, D.R. Soderblom \& R.F.G. Wyse, eds.}
\begin{document}

\maketitle

\begin{abstract} It is well established that activity and rotation
diminishes during the life of sun-like main sequence ($\sim$F7-K2V)
stars. Indeed, the evolution of rotation and activity among these
stars appears to be so deterministic that their rotation/activity
diagnostics are often utilized as estimators of stellar age.  A
primary motivation for the recent interest in improving the ages of
solar-type field dwarfs is in understanding the evolution of debris
disks and planetary systems.  Reliable isochronal age-dating for
field, solar-type main sequence stars is very difficult given the
observational uncertainties and multi-Gyr timescales for significant
structural evolution. Observationally, significant databases of
activity/rotation diagnostics exist for field solar-type field dwarfs
(mainly from chromospheric and X-ray activity surveys). {\it But how
well can we empirically age-date solar-type field stars using
activity/rotation diagnostics?} Here I summarize some recent results
for F7-K2 dwarfs from an analysis by Mamajek \& Hillenbrand (2008),
including an improved ``gyrochronology'' [Period(color, age)]
calibration, improved chromospheric (\rhk) and X-ray (\loglxlbol)
activity vs. rotation (via Rossby number) relations, and a
chromospheric vs. X-ray activity relation that spans four orders of
magnitude in \loglxlbol.  Combining these relations, one can produce
predicted chromospheric and X-ray activity isochrones as a function of
color and age for solar type dwarfs.  

\keywords{Sun: (activity, rotation), stars: (activity, chromospheres,
coronae, fundamental parameters, late-type, low-mass,
rotation), Galaxy: evolution}
\index[subject]{Sun: activity}
\index[subject]{Sun: rotation}
\index[subject]{stars: activity}
\index[subject]{stars: chromospheres}
\index[subject]{stars: coronae}
\index[subject]{stars: fundamental parameters}
\index[subject]{stars: late-type}
\index[subject]{stars: low-mass}
\index[subject]{stars: rotation}
\index[subject]{Galaxy: evolution}
\end{abstract}

\firstsection 

\section{Introduction}\label{intro}

Observational and theoretical studies regarding the evolution of
circumstellar disks and planetary systems have fueled a renewed
interest in assessing how accurately we can determine the ages for
solar-type field dwarfs \citep[][; Meyer, this
volume]{Mamajek08b}. For field stars we can place reasonable
constraints on their effective temperatures, luminosities, and
metallicities from spectroscopic, photometric, and astrometric
measurements. Plotting these observables against theoretical
evolutionary tracks allows us to infer ages and masses. However, for
main sequence solar-type stars, the observational uncertainties in a
star's HRD diagram position can be large enough that they encompass a
large fraction of the star's main sequence lifetime, even for stars
with precise distances and metallicities
\citep[e.g.][]{Nordstrom04, Valenti05, Takeda07}. 
The situation is worse for the hordes of stars lacking trigonometric
parallaxes and metallicity estimates. For this reason, we are
motivated to explore alternative age indicators beyond deriving
individual isochronal ages.

It has been long appreciated that solar-type stars lose angular
momentum via a magnetized wind, spin down, and become less active
during their main sequence phase \citep[e.g.][]{Skumanich72,
Soderblom91}. Here I discuss recent efforts by the author and
collaborators to improve the estimation of ages for solar-type
($\sim$F7-K2) field dwarfs using rotation/activity diagnostics. By
``activity diagnostics'', I will discuss two common examples: the Ca
II H \& K chromospheric activity index \logrphk, and the
X-ray-to-bolometric luminosity ratio
\loglxlbol (= \logrx) in the 0.2-2.4 keV band (ROSAT band). 
For more exhaustive and wavelength-balanced reviews, especially from
the perspective of the Sun's evolution, I refer the reader to
\citet{Gudel07}, \citet{Ayres97}, and \citet{Walter91}.

For the Sun, radiometric dating of the oldest meteorites have
converged on an age within a few Myr of 4.57 Gyr
\citep[e.g.][]{Baker05}. Pleasingly, solar models
which match the observed helioseismological constraints (sound speed
profiles, acoustic modes) can produce the Sun with an age within a few
percent of the meteoritic age
\citep{Houdek08}. For members of nearby young open clusters (e.g. the
Pleiades, Hyades, etc.), detailed modelling of the HR diagram
positions of the high-mass members, and HRD positions and Li-depletion
pattern of the low-mass members, has led to age-dating with claimed
accuracy of $\sim$5-15\% \citep[e.g.][]{deBruijne01,
Barrado04}. Recent results for the small sample of solar-type stars
which have been asteroseismologically observed and modeled are also
yielding age uncertainties of typically $\sim$5-15\%
\citep{Thevenin02, Eggenberger04}, with some claims of even
$\sim$1\%\,
\citep{Mosser08}. While the observational data for these well-studied
clusters and asteroseismological target stars is impeccable, the
accuracy of the inferred cluster ages hinge on the input physics
(e.g. treatment of convection, opacities, etc.) and abundances of the
stellar evolution models, both of which are intimately tied to solar
modeling efforts.  While the Sun provides us with a ``gold'' age
standard ($<$few \% accuracy) and open clusters and
asteroseismological targets provide us ``silver'' age standards
($\sim$15\%), how well can we estimate ages for solar-type field
dwarfs using ``bronze'' indicators like rotation and activity?

\section{Ages from activity and/or rotation}\label{s:rotation}

\subsection{The chromospheric activity-age correlation}

\begin{figure}[htb!]
\begin{center}
\includegraphics[width=2.5in]{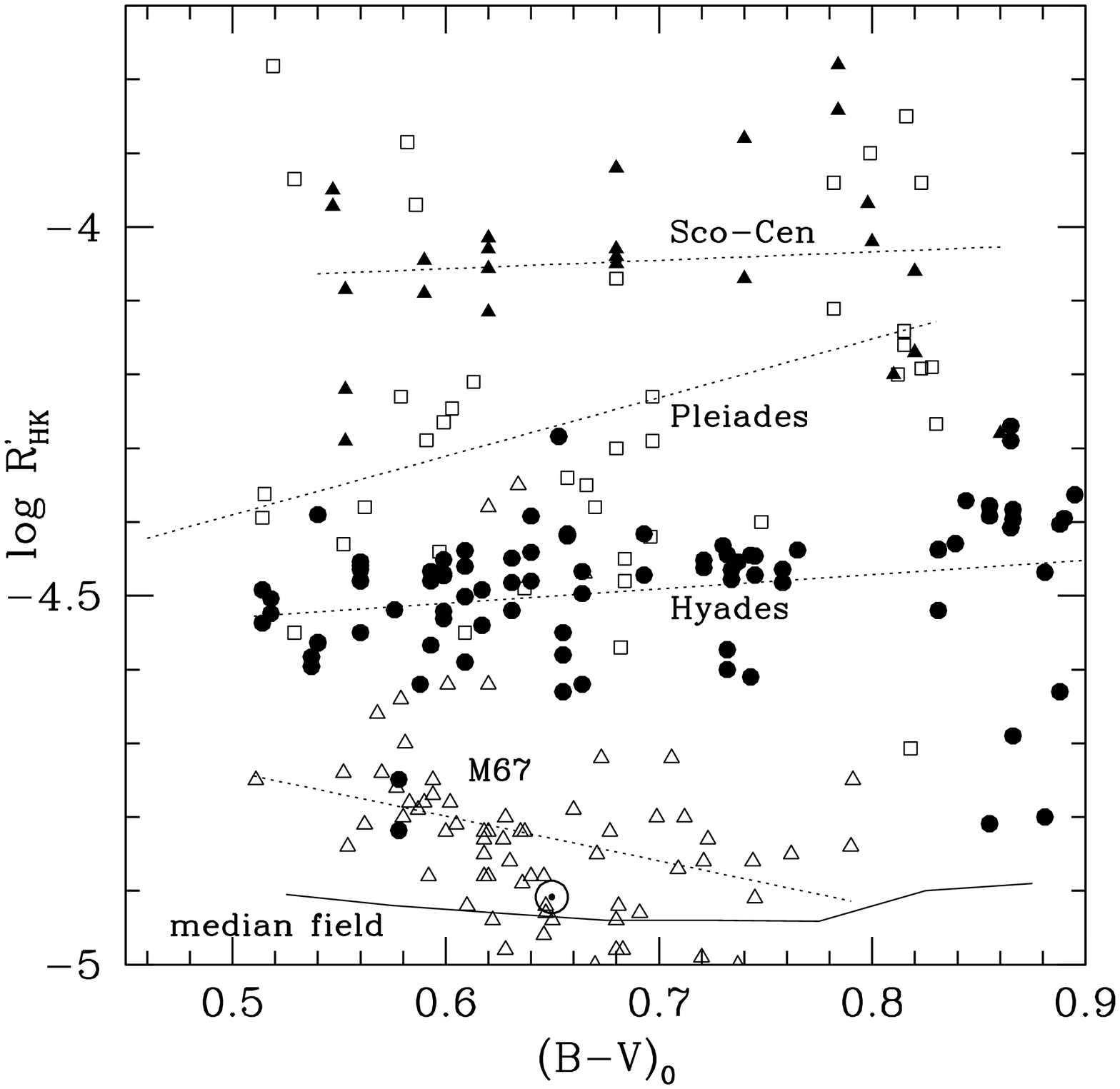}
\includegraphics[width=2.5in]{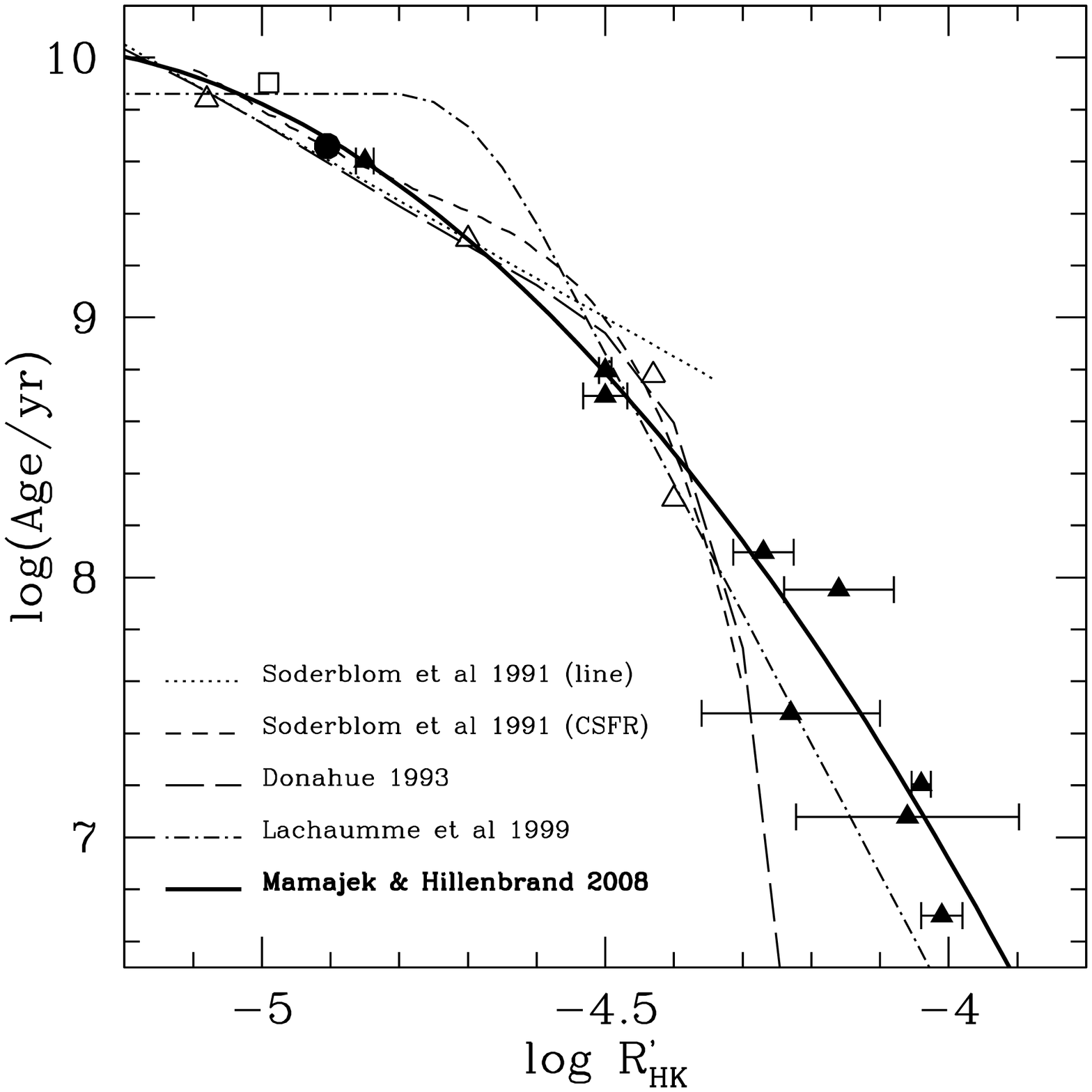}
\caption{{\it Left:} Intrinsic B-V color vs. chromospheric activity index
\logrphk\, for members of age-dated clusters and the Sun 
\citep[from][]{Mamajek08}. Sco-Cen
members are typically $\sim$5-17 Myr old (filled triangles), the
Pleiades are $\sim$130 Myr old (open squares), the Hyades are
$\sim$625 Myr old (filled circles), and M67 is $\sim$4 Gyr old
\citep[open triangles; see][and references therein]{Mamajek08}. The
median \logrphk\, value for field dwarfs as a function of \bvo\, is
plotted. {\it Right:} Color-independent activity-age relationships
from previous studies and \citet{Mamajek08}. The best fit to the
cluster data and a sample of isochronally-dated older field dwarfs is
log\,($\tau$/yr)\, = -38.053 - 17.912\,\logrphk\, -
1.6675\,\logrphk$^2$ (solid line). However cluster and binary
\logrphk\, data suggests that one needs to allow for
color-dependence. This is best taken into account by converting
\logrphk\, $\rightarrow$ rotation $\rightarrow$ age via the Rossby number
and a gyrochronology relation.}
\label{fig:bv_rhk_cluster} 
\end{center} 
\end{figure}

In solar-type stars, the majority of chromospheric and X-ray activity
is believed to be generated as a result of the stellar magnetic
dynamo. The strength of the dynamo and its ability to nonthermally
heat the outer atmospheres of Sun-like stars is ultimately tied to
stellar rotation -- and more specifically -- differential rotation
\citep{Noyes84, Donahue96}. Both activity and rotation among Sun-like
stars are observed to decay with isochronal age \citep[e.g.][]{Wilson63,
Skumanich72, Soderblom91}. As an illustration of this, in
Fig. \ref{fig:bv_rhk_cluster} (left) we plot chromospheric activity
\logrphk\, vs. color for the Sun and members of age-dated clusters
\citep[][]{Mamajek08}. Using our large modern database of activity
and age estimates, we find shortcomings among all previous
activity-age relations.  The new fit to the cluster and field star
activity-age data is shown as a solid line in
Fig. \ref{fig:bv_rhk_cluster} (right):

\begin{equation}
\log\,\tau = -38.053 - 17.912\, \log\,R^{'}_{HK} - 1.6675\, \log\,(R^{'}_{HK})^2
\end{equation}

As is obvious from the leftside figure of
Fig. \ref{fig:bv_rhk_cluster}, there are color-dependent effects which
force us to dismiss a simple activity-age relationship as an
oversimplification. Simply using this activity-age polynomial can
provide age estimates of $\sim$$\pm$0.25 dex or 60\% accuracy
(1$\sigma$; uncertainties come from investigating the scatter in
inferred ages among coeval cluster or binary samples), however
color-dependent systematic effects will be present.
\citet{Mamajek08} were unable to find a simple way
to parameterize age as a function of activity and color which
simultaneously satisfied the available cluster, binary, and field
star datasets. But there is hope in the form of the activity-rotation
correlation \citep[e.g.][]{Noyes84} and the gyrochronology relations
\citep{Barnes07}.

\subsection{Ages from chromospheric activity via rotation}

Theoretical models exist for explaining the decay of rotation speeds
among solar-type stars due to angular momentum loss via magnetized
winds and changes in the moment of inertia of the star
\citep[][]{Kawaler88}. Some of the model parameters are poorly constrained,
(e.g. mass loss, magnetic field geometry), but large rotation period
datasets for clusters can be used to constrain the parameters (Irwin,
this volume). Empirically, however, one can fit a series of simple
curves in color-period-age space. These ``gyrochronology'' curves
introduced by \citet{Barnes07} and improved upon by \citet{Mamajek08}
can be used to derive ages from rotation rates with statistical
accuracy of order $\sim$15\%. For the solar-type dwarfs,
\citet{Mamajek08} fit a gyrochronology relation for period $P$ in days
and age $t$ in Myr:

\begin{equation} 
P(B-V,t) = (0.407\pm0.021)[(B-V)_o -
0.495\pm0.010]^{0.325\pm0.024}~(t/{\rm
Myr})^{0.566\pm0.008}
\end{equation}

Rotation rate can be tied to dynamo strength via the Rossby number
(\ro), which is observationally defined as the rotation period divided
by an estimate of the local convective turnover time just above the
convective-radiative boundary
\citep[$\tau_{c}$; e.g.][]{Noyes84}. Using the best available data
for solar-type dwarfs, \citet{Mamajek08} find the strong correlation
between the Rossby number and chromospheric activity for ``normal''
and ``inactive'' $\sim$F7-K2 dwarfs (\logrphk\, $<$ -4.35) to be:

\begin{equation}
{\rm log}\,R^{'}_{HK}\, = \, -4.522 - 0.337 (R_o - 0.814)
\end{equation}

\begin{figure}[h]
\begin{center}
\includegraphics[width=3in]{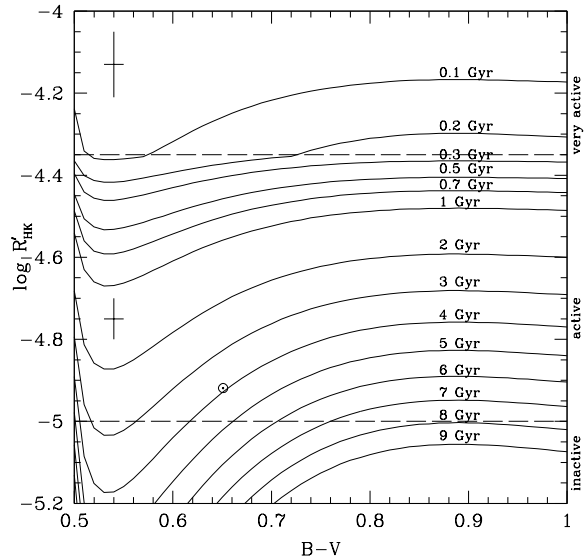}
\caption{Predicted chromospheric activity levels as a function of age
(``gyrochromochrones''), from combining the age-rotation relations
(Eqn. 2.2) with the rotation-activity relations (Eqn. 2.3 for age
$>$100 Myr).  Typical uncertainty bars are shown in the very active
and active regimes, reflecting the r.m.s. in the Rossby
number-activity fits, and typical photometric errors. The behavior of
the gyrochromochrones at the blue end (i.e. the obvious upturn) is not
well-constrained, and is particularly sensitive to the $c$ parameter
in the gyrochronology fits. Figure from \citet{Mamajek08}.
\label{f:gyrochromo}}
\end{center}
\end{figure}

Through applying the conversion activity $\rightarrow$ rotation
$\rightarrow$ age via the Rossby number and gyrochronology, an
analysis of (presumably) coeval stars in resolved binaries and star
clusters suggests that the derived ages have precision of
$\sim$$\pm$0.1-0.2 dex ($\sim$25-50\%; 1$\sigma$). By combining a
gyrochronology relation with the
\logrphk-Rossby number correlation, one can predict chromospheric
activity as a function of color and age for solar-type dwarfs
(``gyrochromochrones''; Fig. \ref{f:gyrochromo}). When combining the
activity vs. rotation and rotation vs. age relations, it becomes
apparent that for a given chromospheric activity level \logrphk, the
late F-type and early G-type stars are systematically younger than the
late G-type and early K-type stars. In the future, it will be prudent
to take into account the effects of metallicity in
Fig. \ref{f:gyrochromo}, and produce isochrones in
mass-metallicity-activity.

\subsection{Coronal X-rays as an age indicator}

It is clear that one can estimate useful stellar ages for solar-type
dwarfs from rotation periods \citep[e.g.][]{Barnes07}, however
rotation periods are currently difficult to measure in large numbers
for all but the most active, starspotted stars \citep[although periods
can be inferred from long-term monitoring of chromospheric
emission;][]{Donahue96}. For these older stars lacking rotation
periods, of order $\sim$10$^{3.5}$ solar-type field dwarfs have
chromospheric activity measurements
\citep[predominantly from large surveys by e.g.][]{Henry96, Wright04,
Gray06} to help us estimate their ages. However, another larger, and
mostly untapped, stellar activity database exists for age estimation:
X-ray fluxes.

Using {\it ROSAT} soft X-ray (0.2-2.4 keV) fluxes,
\citet{Sterzik97} demonstrated that coronal X-ray activity
(\loglxlbol\, = \logrx) scales with chromospheric activity \logrphk\,
over $\sim$4 orders of magnitude in \logrx\, and $\sim$1 order of
magnitude in \logrphk. \citet{Mamajek08} improved the correlation
through including more high- and low-activity stars, and provided
an improved quantification of this correlation for solar-type dwarfs: 

\begin{equation}
{\rm log}\,R^{'}_{HK}\, = (-4.54 \pm 0.01) + (0.289 \pm 0.015)\,({\rm log}\,R_X + 4.92)
\label{eqn:rhk_rx}
\end{equation} 

\noindent with an r.m.s. scatter of 0.06 in \logrphk. The inverse
relation is:

\begin{equation}
{\rm log}\,R_X\,  = (-4.90 \pm 0.04) + (3.46 \pm 0.18)\,({\rm log}\,R^{'}_{HK} + 4.53)
\label{eqn:rx_rhk}
\end{equation} 

\noindent with an r.m.s. of 0.19 dex ($\sim$55\%) in \logrx.  Equation
\ref{eqn:rx_rhk} is statistically consistent with the relation found
by \citet{Sterzik97}, but our uncertainties are $\sim$2$\times$
smaller.

As with the chromospheric activity, their is a strong correlation between
coronal X-ray activity and rotation via the Rossby number:

\begin{equation}
{\rm R_o} = (0.86 \pm 0.02) - (0.79 \pm 0.05)\,(\log R_X + 4.83)
\label{eqn:ro_logrx}
\end{equation} 

This best fit appears to be useful over $\sim$4 orders of magnitude in
\logrx\, (-7 $<$ \logrx\, $<$ -4). The fit is similar to the
linear-log fit quoted by \citet{Hempelmann95}, but is severely at odds
with the oft-cited \logrx\, vs. log \ro\, fit quoted by
\citet{Randich96}. Hence, one can relate X-ray fluxes to rotation periods
via the Rossby number (typically $\sim$0.25 1$\sigma$ accuracy in
\ro), and estimate ages from the periods via a gyrochronology
relation. The typical spread in \logrx\, as a function of age is
$\pm$\,0.4 dex (1$\sigma$) and should be factored into the age
uncertainty. It appears that a few hundred second X-ray snapshot with
an X-ray satellite can be used to predict the multi-decadal average
value of \logrphk\, to within $\pm$\,0.1 (1$\sigma$) accuracy (minimum
age precision $\sim$30-50\%).  A star with X-ray emission similar to
that of the Sun can be seen out to $\sim$15 pc in the ROSAT All-Sky
Survey, \index[subject]{surveys: ROSAT All-Sky Survey} and younger,
more active stars can be seen to larger distances. Using X-ray
emission measured by the ROSAT All-Sky Survey, one should (in
principle) be able to derive useful ages for $\sim$10$^{2-3}$
solar-type field dwarfs in the solar neighborhood.

The very active stars (\logrphk\, $>$ -4.35; \logrx\, $>$ -4.0) have
negligible correlation between rotation and activity (i.e. the
``saturated'' regime). So while setting upper limits to ages might be
fruitful in the saturated regime, quoting exact ages appears not to
be.

\begin{figure}[htb]
\includegraphics[width=2.5in]{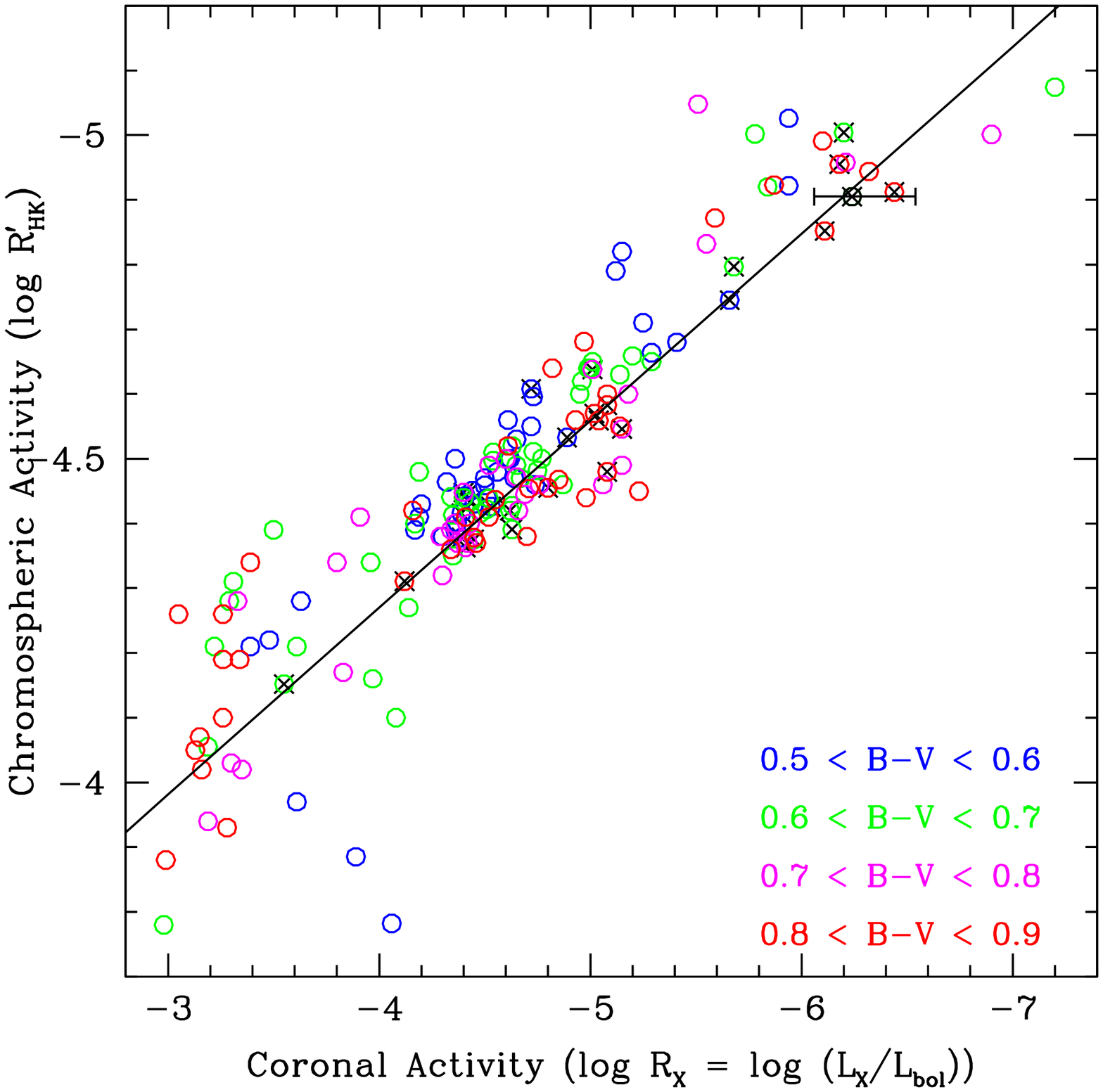}
\includegraphics[width=2.5in]{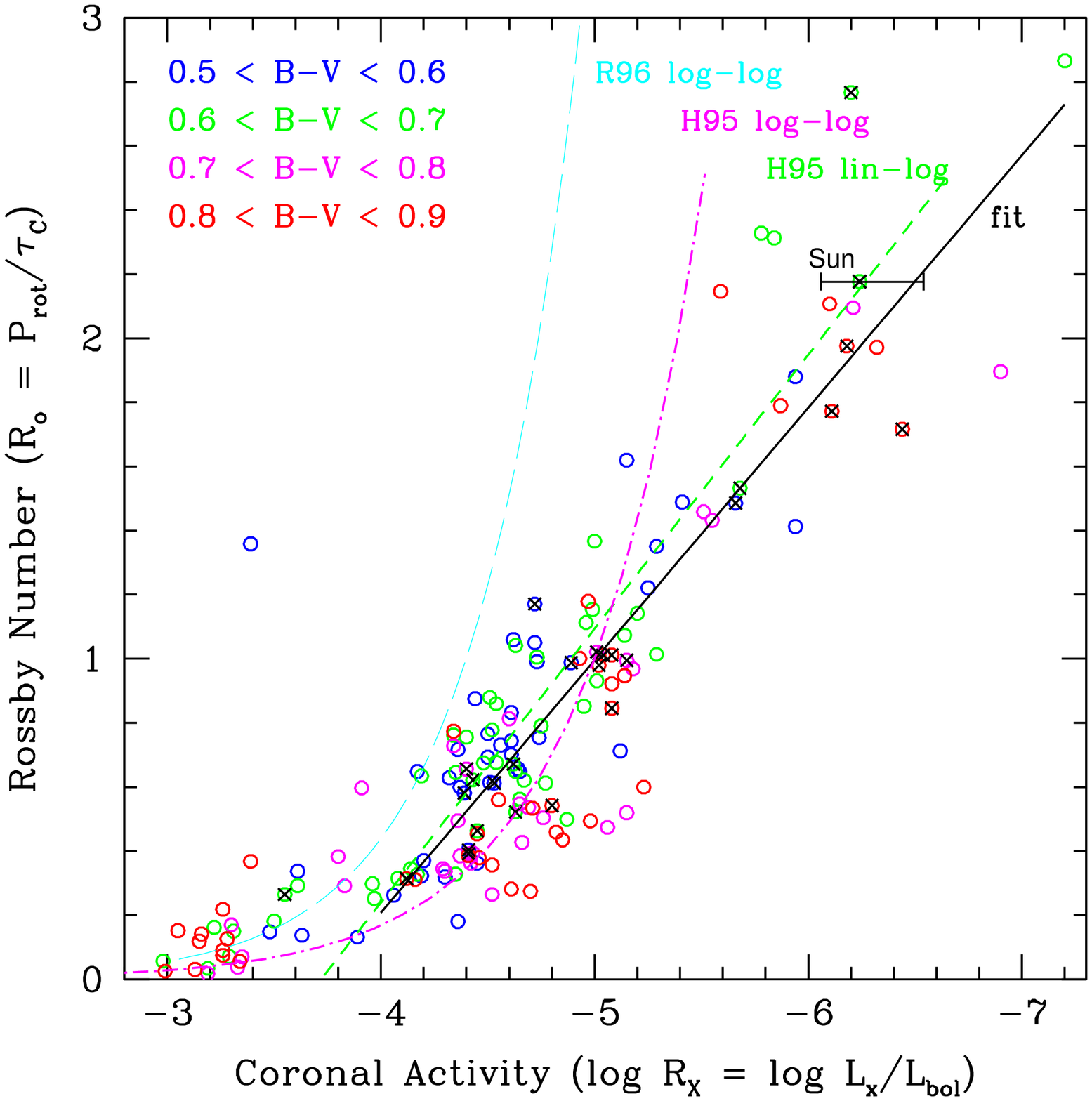}
\caption{{\it Left:} \logrx\, vs. \logrphk\, for solar-type dwarfs with 
known rotation periods and chromospheric and X-ray activity levels
\citep[from][]{Mamajek08}. Stars from \citet{Donahue96} and
\citet{Baliunas96} with well-determined periods also have dark Xs and
conveniently provide an X-ray unbiased sample via the {\it ROSAT}
All-Sky Survey.  Shaded color bins are illustrated in the legend.  The
Solar X-ray and \logrphk\, datum is described in \citet{Mamajek08}.
{\it Right:} \logrx\, vs. Rossby number \ro\, for stars in our sample
of solar-type stars with known rotation periods and chromospheric and
X-ray activity levels \citep[from ][]{Mamajek08}. Donahue-Baliunas
stars with well-determined periods also have dark Xs. Previously
published \rx\, vs. \ro\, fits are drawn: {\it cyan long-dashed line}
is a log-log fit from
\citet{Randich96}, {\it magenta dot-dashed line} is a log-log fit from
\citet{Hempelmann95}, and the {\it green dashed line} is a linear-log
fit from \citet{Hempelmann95}. Our new log-linear fit for stars in the
range -7 $<$ \logrx\, $<$ -4 is the {\it solid dark line}, consistent
with the Hempelmann linear-log relation. Saturated X-ray emission
(\logrx\, $>$ -4) is consistent with \ro\, $<$ 0.5.
\label{f:rhk_logrx}}
\end{figure}

\section{Implications for nearby Sun-like field dwarfs}

What we have constructed are useful empirical relations between
stellar measurements, which when combined, yield a parameter more
difficult to measure: age. In future work, we would like to understand
the physics underlying these empirical relations in terms of how it
constrains stellar dynamo theory \citep[e.g.][]{Montesinos01} and the
evolution of stellar angular momentum \citep[e.g.][]{Kawaler88}. For
the time being, let us use our new and improved rotation-activity-age
tools to see what their implications are.

\begin{figure}[htb]
\begin{center}
\includegraphics[width=3in]{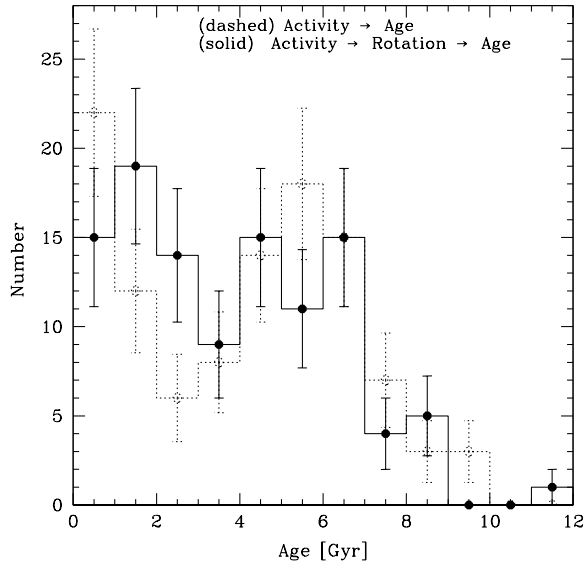}
\caption{Histogram of inferred ages for the nearest 108 solar-type
dwarfs (F7-K2V) within 16 pc \citep[from][]{Mamajek08}. {\it Dashed
histogram} is for ages inferred directly from chromospheric activity
using equation
3. {\it Solid histogram} is for ages derived from converting activity
\logrphk\, to rotation period, then converting rotation period and
color to age via the Rossby number and revised gyro relation.
\label{fig:hist_near}}
\end{center}
\end{figure}

We have already seen that the evolution of activity appears to be
fairly color/mass-dependent among solar-type dwarfs
(Fig. \ref{f:gyrochromo}), contrary to previous studies which employed
a color/mass-independent activity-age correlation. As a first use of
our activity $\rightarrow$ rotation $\rightarrow$ age calibrations, we
constructed a histogram of the chromospheric activity-derived ages for
a volume-limited ($d$ $<$ 16 pc) sample of the nearest 108 solar-type
dwarfs to the Sun\footnote{The ages for these nearest solar-type
dwarfs will be of astrobiological interest for proposed missions
designed to image and take spectra of extrasolar terrestrial planets,
like the TPF and Darwin \citep{Kaltenegger07} and New Worlds Observer
\citep{Cash05}}.  A table of the names, parallaxes, B-V colors,
\logrphk\, values, absolute magnitudes, spectral types, and inferred
ages for the sample stars is given in Table 13 of \citet{Mamajek08}.
We plot the fruits of this effort as a histogram of inferred ages in
Figure \ref{fig:hist_near}. The histogram can not be directly
interpreted as a ``star-formation history'' at this time, as we have
not accounted for the effects of kinematic disk-heating, the loss of
some older stars due to stellar evolution (given the constraint that
the ``dwarf'' stars must lie within 1 mag of the main sequence), and
the effects of metallicity on the sample. As these effects will mostly
conspire to skew our conclusions regarding the old end of the
histogram (i.e. evolved and/or metal-poor and/or high-velocity stars),
we focus on the stars younger than the Sun. First, we note that when a
simple activity $\rightarrow$ age relation is adopted, one sees a
pronounced dip in the age histogram at age $\sim$2-3 Gyr, right in the
region of the ``Vaughan-Preston gap'' \citep{Vaughan80}.  One one
derives age using the recommended activity $\rightarrow$ rotation
$\rightarrow$ age (via the Rossby number and revised gyrochronology
relations), one gets a much flatter age distribution between 0-6 Gyr.
As our color-magnitude selection biases should have negligible impact
on the age distribution of these young to middle-aged dwarfs, it
appears that the histogram is {\it consistent with a more-or-less flat
star-formation history over the past $\sim$5 Gyr or so}. The histogram
is in disagreement with assertions in previous historical studies
\citep[e.g.][ which used an activity-age relation]{Barry88} which
concluded that there has been a recent enhancement of the stellar
birth-rate in the past $\sim$Gyr, which followed a lower birth-rate
$\sim$2-3 Gyr ago. Applying our age-estimation methods to a larger
sample of the nearest solar-type stars out to $\sim$25-40 pc should
place our conclusions on firmer statistical footing. 

These techniques should also be prove valuable for more accurately
assessing the ages of extrasolar planetary systems (Mamajek \&
Slipski, in prep.) and dusty debris disk systems (Hillenbrand et al., in
prep.). 

\begin{acknowledgements} The author collaborated with Lynne
Hillenbrand on much of the material presented for this talk (presented
in Mamajek \& Hillenbrand 2008), and acknowledges helpful
conversations with D. Soderblom, J. Stauffer, and M.~R. Meyer.
\end{acknowledgements}


\begin{discussion}

\discuss{D. Soderblom}{First, maybe it's a coincidence, but the line
for Rossby number $=1$ runs right down the middle of the
"Vaughan-Preston gap," and that may explain the build-up of stars at
$\log R^\prime_{\rm HK} = -4.5$. Second, have you compared activity
ages to isochrones ages? When I do I see zero correlation.}

\discuss{E. Mamajek}{Rossby number $=1$ depends on the choice of
convective overturn time, which differs by a factor of a few among the
models. I adopted those from Noyes et al. (1984). The scatter between
ages from activity or gyrochronology and those from isochrones
(Valenti \& Fischer 2005) is large; however, Valenti has suggested
that Takeda et al. (2007) ages are to be preferred, but I have not yet
compared to those.}

\discuss{F. Walter}{The Rossby number is a convenient way to sweep a
lot of our ignorance about convection under the rug (or at least into
a single parameter). How well do we really understand convective
turnover timescales in convective stars, and how might this
uncertainty affect the details of your activity-age relations?}

\discuss{E. Mamajek}{Modelers have evaluated convective turnover times
differently -- primarily at different depths with respect to the base
of the convection zone -- but there appears to broad agreement in the
{\it relative} turnover times for main sequence stars as a function of
mass. Combining the theoretical turnover times with the observed
rotation and activity data shows that a strong correlation between
Rossby number and activity exists over a wide range of parameter space
for solar-type dwarfs when one uses MLT models with $\alpha$ = 1.9
(see Noyes et al. 1984 and Montesinos et al. 2001). Deriving rough
ages for solar-type dwarfs using activity is then supported by two
empirical correlations: the activity- rotation relation (via the
Rossby number) and the rotation-age relation (gyrochronology; see talk
by S. Barnes).}

\end{discussion}


\begin{thebibliography}{}

\bibitem[Ayres(1997)]{Ayres97} Ayres, T.~R.\ 1997, \jgr, 102, 1641


\bibitem[Baker et al.(2005)]{Baker05} Baker, J., Bizzarro, M., Wittig,
N., Connelly, J., \& Haack, H.\ 2005, \nat, 436, 1127

\bibitem[Baliunas et al.(1996)]{Baliunas96} Baliunas, S., Sokoloff,
D., \& Soon, W.\ 1996, \apjl, 457, L99

\bibitem[Barnes(2007)]{Barnes07} Barnes, S.~A.\ 2007, \apj, 669, 
1167 

\bibitem[Barrado y Navascu\'{e}s, et al.(2004)]{Barrado04} Barrado y
Navascu\'{e}s, D., Stauffer, J.~R., \& Jayawardhana, R. 2004, \apj,
614, 386

\bibitem[Barry(1988)]{Barry88} Barry, D.~C.\ 1988, \apj, 334, 
436 

\bibitem[de Bruijne et al.(2001)]{deBruijne01} de Bruijne, J.~H.~J.,
Hoogerwerf, R., \& de Zeeuw, P.~T.\ 2001, \aap, 367, 111

\bibitem[Cash et al.(2005)]{Cash05} Cash, W., Kasdin, J., 
Seager, S., \& Arenberg, J.\ 2005, \procspie, 5899, 274 

\bibitem[Donahue et al.(1996)]{Donahue96} Donahue, R.~A., Saar,
S.~H., \& Baliunas, S.~L.\ 1996, \apj, 466, 384

\bibitem[Eggenberger et al.(2004)]{Eggenberger04} Eggenberger, P.,
et al.\ 2004, \aap, 417, 235

\bibitem[Gray et al.(2006)]{Gray06} Gray, R.~O., et al.\ 2006, \aj,
132, 161

\bibitem[G{\"u}del(2007)]{Gudel07} G{\"u}del, M.\ 2007, Living Reviews
in Solar Physics, 4, 3

\bibitem[Hempelmann et al.(1995)]{Hempelmann95} Hempelmann, A., et
al.\ 1995, \aap, 294, 515

\bibitem[Henry et al.(1996)]{Henry96} Henry, T.~J., Soderblom, 
D.~R., Donahue, R.~A., \& Baliunas, S.~L.\ 1996, \aj, 111, 439

\bibitem[Houdek \& Gough(2008)]{Houdek08} Houdek, G., \& Gough, D.~O.\
2008, IAU Symposium, 252, 149

\bibitem[Kaltenegger et al.(2007)]{Kaltenegger07} Kaltenegger, L., 
Traub, W.~A., \& Jucks, K.~W.\ 2007, \apj, 658, 598 

\bibitem[Kawaler(1988)]{Kawaler88} Kawaler, S.~D.\ 1988, \apj, 
333, 236 

\bibitem[Mamajek et al.(2008)]{Mamajek08b} Mamajek, E.~E.,
Barrado y Navascu{\'e}s, D., Randich, S., Jensen, E.~L.~N., Young,
P.~A., Miglio, A., \& Barnes, S.~A.\ 2008, 14th Cambridge Workshop on
Cool Stars, Stellar Systems, and the Sun, 384, 374

\bibitem[Mamajek \& Hillenbrand(2008)]{Mamajek08} Mamajek, E.~E., \&
Hillenbrand, L.~A.\ 2008, \apj, 687, 1264

\bibitem[Montesinos et al.(2001)]{Montesinos01} Montesinos, B.,
Thomas, J.~H., Ventura, P., \& Mazzitelli, I.\ 2001, \mnras, 326, 877

\bibitem[Mosser et al.(2008)]{Mosser08} Mosser, B., et al.\ 2008,
\aap, 488, 635

\bibitem[Nordstr{\"o}m et al.(2004)]{Nordstrom04} Nordstr{\"o}m, 
B., et al.\ 2004, \aap, 418, 989 

\bibitem[Noyes et al.(1984)]{Noyes84} Noyes, R.~W., et al.\ 1984,
\apj, 279, 763

\bibitem[Randich et al.(1996)]{Randich96} Randich, S., Schmitt,
J.~H.~M.~M., Prosser, C.~F., \& Stauffer, J.~R.\ 1996, \aap, 305, 785

\bibitem[Skumanich(1972)]{Skumanich72} Skumanich, A.\ 1972, \apj, 171,
565

\bibitem[Soderblom et al.(1991)]{Soderblom91} Soderblom, D.~R.,
Duncan, D.~K., \& Johnson, D.~R.~H.\ 1991, \apj, 375, 722

\bibitem[Sterzik \& Schmitt(1997)]{Sterzik97} Sterzik, M.~F., \&
Schmitt, J.~H.~M.~M.\ 1997, \aj, 114, 1673

\bibitem[Takeda et al.(2007)]{Takeda07} Takeda, G., et al.\ 2007,
\apjs, 168, 297

\bibitem[Th{\'e}venin et al.(2002)]{Thevenin02} Th{\'e}venin, F.,
et al.\ 2002, \aap, 392, L9

\bibitem[Valenti \& Fischer(2005)]{Valenti05} Valenti,
J.~A., \& Fischer, D.~A.\ 2005, \apjs, 159, 141 (VF05)

\bibitem[Vaughan \& Preston(1980)]{Vaughan80} Vaughan, A.~H., \&
Preston, G.~W.\ 1980, \pasp, 92, 385


\bibitem[Walter \& Barry(1991)]{Walter91} Walter, F.~M., \& Barry,
D.~C.\ 1991, The Sun in Time, 633

\bibitem[Wilson(1963)]{Wilson63} Wilson, O.~C.\ 1963, \apj, 138, 832

\bibitem[Wright et al.(2004)]{Wright04} Wright, J.~T., Marcy, G.~W.,
Butler, R.~P., \& Vogt, S.~S.\ 2004, \apjs, 152, 261

\end{thebibliography}
\end{document}